# T1rho Fractional-order Relaxation of Human Articular Cartilage*

Lixian Zou, Haifeng Wang, Yanjie Zhu, Yuanyuan Liu, Jing Cheng, Sen Jia, Caiyun Shi, Shi Su, Xin Liu, Hairong Zheng, and Dong Liang, *Senior Member, IEEE*

*Abstract*—T1rho imaging is a promising non-invasive diagnostic tool for early detection of articular cartilage degeneration. A mono-exponential model is normally used to describe the T1rho relaxation process. However, mono-exponentials may not adequately to describe NMR relaxation in complex, heterogeneous, and anisotropic materials, such as articular cartilage. Fractional-order models have been used successfully to describe complex relaxation phenomena in the laboratory frame in cartilage matrix components. In this paper, we develop a time-fractional order (T-FACT) model for T1rho fitting in human articular cartilage. Representative results demonstrate that the proposed method is able to fit the experimental data with smaller root mean squared error than the one from conventional mono-exponential relaxation model in human articular cartilage.

## I. INTRODUCTION

Osteoarthritis (OA) is a degenerative disease of articular cartilage [1]. The loss of proteoglycans and other macromolecules in the extracellular matrix among the early stage in OA will develop to cartilage thinning and cause patient pain. Conventional morphological MRI techniques are used for the characterization of advanced cartilage lesions. However, it is less sensitive to assess early macromolecular changes at the early stages of OA [2]. The early diagnosis of OA is increasingly important to allow early-intervention therapies.

T1rho imaging [3,4] is a novel promising noninvasive quantitative technology for tissue characterization similar to T1 (spin-lattice relaxation), T2 (spin-spin relaxation) and MT (magnetization transfer) imaging. T1rho is sensitive to the slow motion interactions between motion-restricted water molecules and their local environment. It plays a key role in visualizing macromolecular changes for many biomedical applications. For example, it has demonstrated its potential in characterizing the earliest changes in articular cartilage degeneration [3,4] and in detecting disc degeneration [4,5]. In contrast to T1 or T2 relaxation, T1ρ relaxation characterizes magnetization decay under the application of the spin-lock radiofrequency (RF) pulse [6]. It is able to measure the low frequency processes via applied RF pulse at any currently available clinical MRI field strengths. To measure T1rho relaxation, a 90° pulse flipped the spin magnetization into the transverse plane firstly. Then the so-called spin-lock (SL) pulse with a long, low powered RF pulse is applied parallel to the magnetization. The transverse magnetization decays with a time constant T1ρ around the applied spin lock field. The relaxation affected by spin-lock field in the rotating frame is similar to the longitudinal relaxation around the B0 field in the laboratory frame. As the frequency of the spin-locking pulse is zero, T1ρ relaxation becomes T2 relaxation. T1ρ approaches T1 while frequency of the spin-locking pulse approaches the Larmor frequency. It takes a long scan time as multiple images with different spin-lock times (TSLs) are needed in quantitative 3D T1rho mapping. Several ways to effectively reduce the acquisition time of T1rho mapping are proposed including reducing the numbers of TSLs or fast imaging techniques by acquiring highly undersampling k space data then reconstructed using compress sensing [7-9].

T1rho relaxation process is normally described by a mono-exponential model. However, one-compartment model may not adequately to describe more complex tissues where different proton compartments exist and interact with each other. Thus, two-compartment model is proposed to describe the free water proton and the constraint water proton as well as the interaction between them [10-12]. But there are many anomalous cases being observed such as or power-law behavior [13-16] that bi- or multi- exponential model becomes inadequate to describe NMR relaxation. Fractional order model, thus, is more flexible to describe the dynamics of complex phenomenon including the anomalous NMR relaxation phenomenon. These models have been used successfully to describe complex relaxation phenomena in the laboratory frame in cartilage matrix components and native cartilage [14]. Thus, it is reasonable to consider such models for T1rho relaxation in articular cartilage. In this work, we develop a time-fractional order (T-FACT) model and evaluate the feasibility of the model for T1rho relaxation fitting in human articular cartilage.

## II. THEORY AND METHODS

Fractional calculus is aimed to improve the power of clinical diagnosis through improved modeling. It has been successfully used to extend the classical Bloch equations and

*Resrach partially supported by the National Natural Science Foundation of China (No. 81830056, No. 61471350, No. 61871373, and No. 81729003), National Key R&D Program of China (No. 2016YFC0100100), Natural Science Foundation of Guangdong Province (No. 2017A050501026, and No. 2018A0303130132), Guangdong Provincial Key Laboratory of Magnetic Resonance and Multimodality Imaging (No. 2014B030301013), Shenzhen Key Laboratory for MRI (No. CXB201104220028A), Shenzhen Key Laboratory of Ultrasound Imaging and Therapy (No. ZDSYS20180206180631473), and Innovation and Technology Commission of the government of Hong Kong SAR (No. MRP/001/18X).

L Zou, H Wang, Y Zhu, Y Liu, J Cheng, S Jia, C Shi, S Su, X Liu and H Zheng are with Paul C. Lauterbur Research Centre for Biomedical Imaging, Shenzhen Institutes of Advanced Technology, Chinese Academy of Sciences, Shenzhen, Guangdong, China (e-mail: { lx.zou, hf.wang1, yj.zhu, liuyy, jing.cheng, sen,jia, cy.shi, shi.su, xin.liu, hr.zheng}@siat.ac.cn)

D Liang is with Research Center and Medical AI, and Paul C. Lauterbur Research Centre for Biomedical Imaging, Shenzhen Institutes of Advanced Technology, Chinese Academy of Sciences, Shenzhen, Guangdong, China (corresponding author to provide phone: 86-0755-86392274; e-mail: dong.liang @ siat.ac.cn).

has been proposed to fit the experimental data with an accuracy that is not achievable with the classical mono-exponential model under the laboratory frame in the application of brain and bovine cartilage [14,16], where the fractional derivatives are expressed as Caputo or Riemann-Liouville fractional derivative operators. The definition and properties of the fractional derivative are given in Appendix session.

Similar to T1 and T2 relaxation, the process of spin-lattice relaxation in the rotating frame can be described as time-fractional order relaxation that extends the original relaxation equation.

*A. Time-fractional Order T1rho Fitting Model*

Here, we adopt time-fractional order model (T-FACT) developed by Magin et al [13, 16] to express the process of T1rho relaxation. Thus, the T-FACT model used to fit voxel-wise image intensities with different TSLs is described as

$$S_n = S_0 \cdot E_\alpha \left( \frac{-TSL_n^\alpha}{T'_{1rho}} \right),$$

where $E_\alpha(t)$ is the single-parameter Mittag-Leffler function [13-16], $T'_{1rho} = \tau_1^{\alpha-1} T_{1rho}$, $\tau_1$ is fractional time constant to maintain a consistent set of units, and α is the fractional-order in time. The definition and properties of the Mittag-Leffler function are given in Appendix session. In the case of α=1, the Mittag-Leffler function corresponds to the conventional mono-exponential relaxation process.

*B. Root Mean Square Error*

The data sets were fitted to the mono-exponential (MONO) model and the T-FACT model, respectively. $RMSE_{Model} = \sqrt{\frac{\sum_n (S_{Model}(TSLn) - S_{Acquired}(TSLn))^2}{N}}$ was performed to compare the two relaxation models voxel by voxel, where N the number of TSLs, $S_{Model}(TSLn)$ the intensity of $n$ th TSL from $Model$ (means MONO or T-FACT model) and $S_{Acquired}(TSLn)$ the acquired intensity of $n$th TSL.

*C. Data Acquisition*

Data sets were acquired from a Philips Achieva 3.0TX scanner (Philips Healthcare, Best, the Netherlands) with an eight channel T/R knee coil (Invivo Corp, Gainesville, USA).

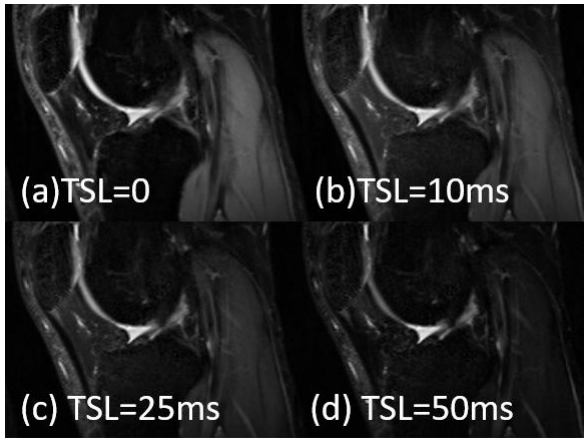

Figure 1. Representative T1rho weighted images acquired at TSL 0, 10, 25, 50 ms, repectively

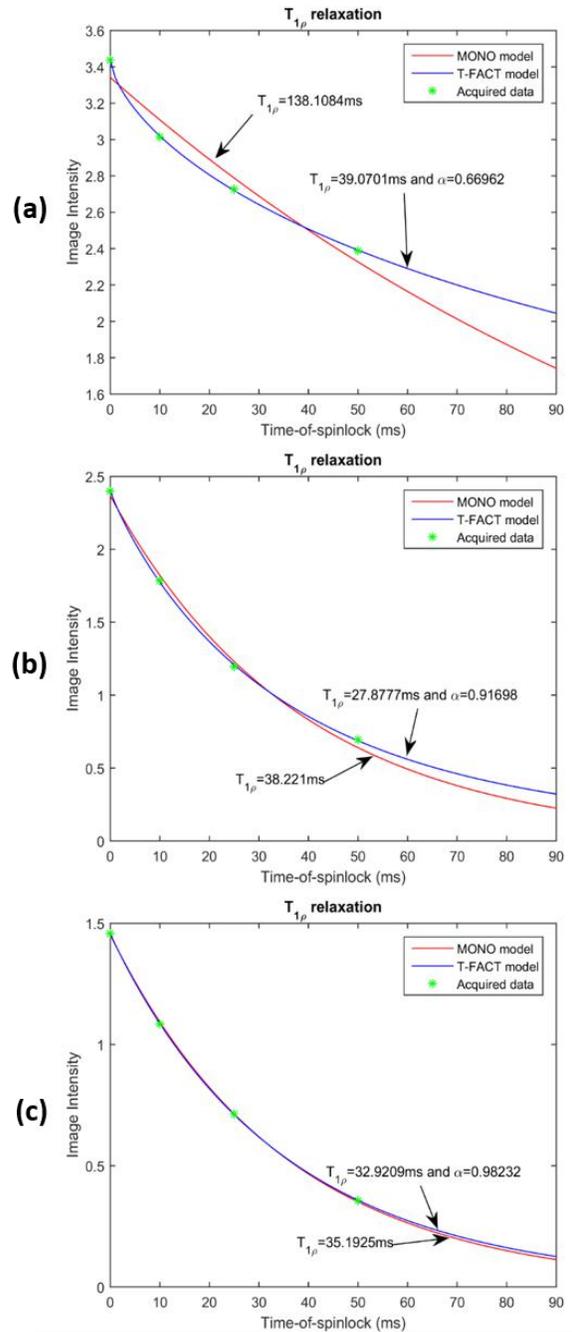

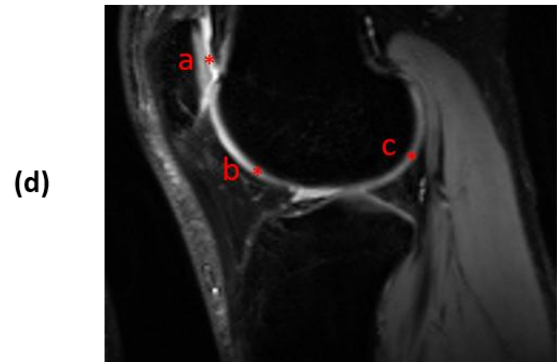

Figure 2. The comparison of fitting results (a,b,c) using time-fractional order (T-FACT) and mono-exponential (MONO) models on three voxels in the articular cartilage shown in (d). (d) T1rho-weigthed image at TSL 0ms.

Table 1 $T_{1\rho}$ from two relaxation models and their average RMSE

| ROI | MONO model | T-FACT model | | Average RMSE | |
|---|---|---|---|---|---|
| | $T_{1\rho}$(ms) | $T_{1\rho}$(ms) | α | MONO | T-FACT |
| 1 | 140.69 ± 97.14 | 63.71 ± 28.29 | 0.8160 | 19.82% | 6.79% |
| 2 | 39.84 ± 7.63 | 32.39 ± 6.49 | 0.9446 | 12.68% | 4.53% |
| 3 | 37.48 ± 3.93 | 33.33 ± 6.81 | 0.9647 | 12.24% | 4.65% |

Sagittal knee scan of one subject was conducted under the approval of the Institutional Review. A T1rho-prepared fast spin echo with fat suppression was used for 39 slices data scanning. The acquisition parameters include: resolution 1 x 1 x 3 mm³, TR/TE = 2200/23 ms, spin-lock frequency 500Hz, and TSL=[0 10 25 50] ms.

## III. RESULTS

In the all slices of the subject, the T-FACT model shows improved fitting of the acquired data compared to the mono-exponential model. Figure 1 shows typical T1rho-prepared images on a representative slice in the articular cartilage of the subject with different TSLs. Three voxels were selected as shown on Figure 2(d) to demonstrate the T1rho relaxation curves (Figure 2) using the two relaxations model. Three regions of interesting (ROI) were delineated surround the three selected voxels. T1rho, the fractional order, and average root mean square error (RMSE) of the three ROIs were listed in Table 1. Average RMSEs show that the T-FACT model fit the relaxation much better than the MONO model. Figure 3 shows the T1rho maps, α map, and RMSE maps of the representative slice of the articular cartilage using two fitting models. And complexity of the tissue is depicted in the α map. Articular cartilage of the patella show a sharp deviation from the mono-exponential model, but follow T-FACT better. The proposed model is able to fit the experimental data with smaller root mean squared error than the classical mono-exponential relaxation model as shown in Figure 3.

## IV. DISCUSSION AND CONCLUSION

Anomalous relaxation illustrated in the first plots on Figure 2(a) leads to the sharp deviation of acquired data from the mono-exponential decay, which results to abnormal T1rho values as shown on the Figure 3. One possible reason could be articular cartilage is no longer satisfy the theoretical assumptions underlying the classical Bloch equations due to the complex environment including synovial fluid and ligament. The proposed T-FACT model widens the scope of application by extending the classical equations and dispenses with other assumption such as compartmentalized signal hypothesis in bi-exponential model. Otherwise, the fractional order α in T-FACT model may capture information for structural complexity inside or outside the human articular cartilage.

The results show the proposed method can better represent the T1rho relaxation in human articular cartilage. In the future, further investigation is needed to understand the contribution of fractional order α and its potential clinical explanation.

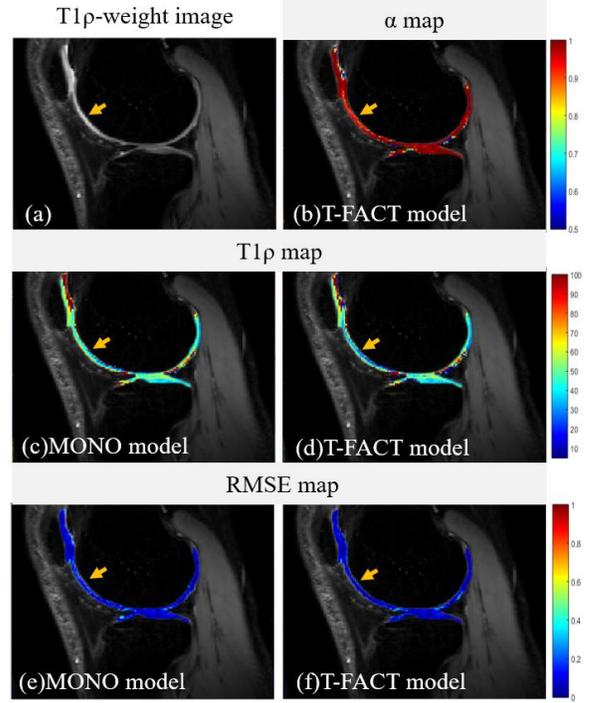

Figure 3. A representative slice in the articular cartilage demonstrates: T1rho–weighted image (a), T1rho maps (c, d) and RMSE maps (e, f) from two relaxation models of MONO and T-FACT, and the α map (b) from T-FACT model. Articular cartilage of the patella show a sharp deviation from the mono-exponential model, but follows T-FACT better. Yellow arrows show RMSE with T-FACT model is lower than that with MONO model.

## APPENDIX

There are various kinds of definitions for fractional derivatives. In this appendix, we introduce the Caputo fractional derivative and some of their properties used in this work (see also [13-16]).

### A. Caputo fractional derivative

Given $\alpha > 0$ with $n-1 < \alpha < n$ for integer values of n, assume that $f(t)$ is a suitable real function of $t > 0$, the Riemann-Liouville integral order of $\alpha$ has the form

$$I^\alpha f(t) = \frac{1}{\Gamma(\alpha)} \int_0^t (t-\tau)^{\alpha-1} f(\tau) d\tau, \quad (1)$$

Where

$$\Gamma(z) = \int_0^\infty e^{-u} u^{z-1} du, \quad (2)$$

is the Gamma function.

Then the Caputo fractional derivative is defined as

$$^C_0 D_t^\alpha f(t) = I^{n-\alpha}(D^n f)(t), t > 0 \quad (3)$$

Since the order α of the derivative is restrict to $0 < \alpha < 1$, the definition can simplify to

$$^C_0 D_t^\alpha f(t) = -\frac{1}{\Gamma(1-\alpha)} \int_0^t (t-\tau)^{-\alpha} \frac{df(\tau)}{d\tau} d\tau, \quad (4)$$

## B. Mittag-Leffler function

Mittag-Leffler function is a generalization of the simple exponential function. It often appears in applications of fractional calculus.

### 1) Single parameter Mittag-Leffler function

The single parameter Mittag-Leffler function defines as

$$E_\alpha(z) = \sum_{k=0}^{\infty} \frac{z^k}{\Gamma(\alpha k + 1)}, \quad \alpha > 0 \quad (5)$$

In the case of $\alpha = 1$, the Mittag-Leffler function has the property $E_1(z) = e^z$, leading to the conventional exponential relaxation processes.

Another property used in this work is

$${}_0^C D_t^\alpha E_\alpha(\lambda z^\alpha) = \lambda E_\alpha(\lambda z^\alpha), \quad (6)$$

### 1) Two-parameter Mittag-Leffler function

The two-parameter Mittag-Leffler function has the form

$$E_{\alpha,\beta}(z) = \sum_{k=0}^{\infty} \frac{z^k}{\Gamma(\alpha k + \beta)}, \quad \alpha, \beta > 0 \quad (7)$$

When $\beta = 1$, it leads to the single parameter Mittag-Leffler function. It has the property

$$E_{1,2}(z) = \frac{e^z - 1}{z}, \quad (8)$$

ACKNOWLEDGMENT

The authors will thank Dr. Weitian Chen for sharing the data.

REFERENCES

[1] R. Gnannt, "MR imaging of the postoperative knee," *J Magn Reson Imaging*, vol. 34, no. 5, pp. 1007-1021, 2011
[2] H. Shao, "Magic angle effect plays a major role in both T1rho and T2 relaxation in articular cartilage," *Osteoarthritis Cartilage*, vol. 25, no. 12, pp. 2022-2030, 2017
[3] W.R. Witschey, "T1rho MRI quantification of arthroscopically confirmed cartilage degeneration," *Magn Reson Med*, vol. 63, no. 5, pp. 1376-1382, 2010..
[4] Y. J. Wang, "T1ρ magnetic resonance: basic physics principles and applications in knee and intervertebral disc imaging," *Quantitative imaging in medicine and surgery*, vol. 5, pp. 858-885, 2015.
[5] C.P.L. Paul, "Quantitative MRI in early intervertebral disc degeneration: T1rho correlates better than T2 and ADC with biomechanics, histology and matrix content," *PLoS One*, vol. 13, no. 1, pp. e0191442, 2018.
[6] R.E. Sepponen, "A method for T1rho imaging," *Journal of Computer Assisted Tomography*, 1985.
[7] Y. Zhu, "PANDA-T1rho: Integrating principal component analysis and dictionary learning for fast T1rho mapping," *Magn Reson Med*, vol. 73, no. 1, pp. 263-272, 2015.
[8] Y. Zhou, "Accelerating T1rho cartilage imaging using compressed sensing with iterative locally adapted support detection and JSENSE," *Magn Reson Med*, vol. 75, no. 4, pp. 1617-1629, 2016.
[9] M.V.W. Zibetti, "Accelerating 3D-T1rho mapping of cartilage using compressed sensing with different sparse and low rank models," *Magn Reson Med*, vol. 80, no. 4, pp. 1475-1491, 2018
[10] R.G. Menon, "Bi-exponential 3D-T1rho mapping of whole brain at 3 T," *Sci Rep*, vol. 8, no. 1, pp. 1176, 2018.
[11] J.Q. Han, "Accuracy and reproducibility of T1rho mapping sequences," *Journal of Cardiovascular Magnetic Resonance*, vol.17, pp. 22, 2015.
[12] J. Yuan, "Observation of bi-exponential T(1rho) relaxation of in-vivo rat muscles at 3T," *Acta Radiol*, vol. 53, no. 6, pp. 675-681, 2012.
[13] R. L. Magin, "Solving the fractional order Bloch equation," *Concepts Magn. Reson.*, vol. 34, pp. 16–23, 2009.
[14] H. Wang, "Application of Time-Fractional Order Bloch Equation in Magnetic Resonance Fingerprinting" *arXiv preprint*, arXiv:1904.02332, 2019.
[15] R. L. Magin, "Anomalous NMR relaxation in cartilage matrix components and native cartilage: Fractional-order models," *Journal of Magnetic Resonance*, vol. 210, pp. 184-191, 2011.
[16] S. Qin, "Characterization of anomalous relaxation using the time-fractional Bloch equation and multiple echo T2*-weighted magnetic resonance imaging at 7 T," *Magn. Reson. Med.*, vol. 77, pp. 1485-1494, 2017.